\documentclass[manuscript]{acmart}
\usepackage{graphicx}
\usepackage{color,soul, amsmath, url}
\usepackage{subcaption}
\usepackage{tabularx}

\AtBeginDocument{%
  \providecommand\BibTeX{{%
    \normalfont B\kern-0.5em{\scshape i\kern-0.25em b}\kern-0.8em\TeX}}}

\begin{document}

\title{Open Image Content Disarm And Reconstruction}

\author{Eli Belkind}
\email{eli.belkind@msmail.ariel.ac.il}

\author{Ran Dubin}
\email{rand@ariel.ac.il}
\orcid{0000-0002-2055-2211}

\author{Amit Dvir}
\email{amitdv@ariel.ac.il}
\orcid{0000-0002-3670-0784}

\affiliation{%
  \institution{\\Department of Computer Science, Ariel Cyber Innovation Center (ACIC)}
  \country{Ariel University, Ariel, Israel}
}

\renewcommand{\shortauthors}{Belkind et al.}

\begin{abstract}
With the advance in malware technology, attackers create new ways to hide their malicious code from antivirus services. One way to obfuscate an attack is to use common files as cover to hide the malicious scripts, so the malware will look like a legitimate file.
Although cutting-edge Artificial Intelligence and content signature exist, evasive malware successfully bypasses next-generation malware detection using advanced methods like steganography.
Some of the files commonly used to hide malware are image files (e.g., JPEG). In addition, some malware use steganography to hide malicious scripts or sensitive data in images. Steganography in images is difficult to detect even with specialized tools.
Image-based attacks try to attack the user's device using malicious payloads or utilize image steganography to hide sensitive data inside legitimate images and leak it outside the user's device.
Therefore in this paper, we present a novel Image Content Disarm and Reconstruction (ICDR). Our ICDR system removes potential malware, with a zero trust approach, while maintaining high image quality and file usability. By extracting the image data, removing it from the rest of the file, and manipulating the image pixels, it is possible to disable or remove the hidden malware inside the file\footnote{The code and part of the dataset and some results (due to space limitations) can be found in https://github.com/ArielCyber/ICDR}. 
\end{abstract}

\maketitle

\section{Introduction}
With the widespread use of the Internet, there is more interest and development in cyber attacks. Cyber attackers attempt to exploit vulnerabilities of widely-used sites, apps, and files to harm Internet users while remaining undetected. For example, the extreme popularity of social media has brought a new hype in sharing images between users. This popularity of image sharing can be exploited by camouflaging malware inside normal-looking image files and using social media to distribute and potentially attack users through distribution on social media platforms, infected websites, and messaging applications  \cite{Karlo2021malware,dong2020ImageDetox}. Potentially, victims of these attacks can download the files, and with an external program or by simply opening the file, the malicious script runs and attacks the user \cite{Karlo2021malware, MS05009, jellyfish}. Alternatively, an attacker can upload an infected image with malicious PHP code to social media, resulting in the attacker gaining access to the servers of the platform \cite{owaspFileUpload}. In order to deal with these types of attacks, users will need a service that can scan and detect downloaded malicious images. However, the problem with this method is that implementing such a service with robust functionality is difficult. Therefore, the service provider/organization that receives the files should scan all incoming images, while the end-users who consume the images should scan the incoming images and ensure they are safe to use and do not contain any hidden (steganography) or malicious content. Meaning, this type of service is time and resource consuming \cite{aviad2020maljpeg,nandhini2020image}. A traditional file/image inspection pipeline consists of multi antivirus detection (static analysis/machine learning) engines and malware sandbox detonation (dynamic run time analysis) \cite{serpanos2021sisyfos}. At the same time, both antivirus and sandbox solutions are detection mechanisms that rely in the end on file detection \cite{MarkBaker}, meaning file trust. However, it is still possible that evasive malware will not be detected. Recent antivirus detection tests show that the best solution achieves a 96.3\% detection rate \cite{avcompare} which is insufficient. Therefore, a zero-trust prevention layer is needed. A Content Disarm and Reconstruct (CDR) system is zero-trust prevention methodology. CDR extracts the original (benign) content, disarming all potential threat attack vectors and reconstructing the file so it is fully functional and secure \cite{wiseman2017content, dubin_CDR}. CDR solutions require an understanding of the file's structure and its various components to separate malicious and benign elements. This understanding allows the CDR to "clean" the file without corrupting it. Therefore, a CDR method can be very robust, removing zero-day attacks that can remain undetected by conventional anti-virus services.

The main advantages of CDR technology are running speed, the ability to be added to Anti-Virus/sandbox detection steps, and increasing safety by using a zero-trust approach for files. However, some inherent limitation exists with CDR, such as: 
(1) Decrease in the file usability caused by the removal of active code and, functionality intended for legitimate use; (2) Problematic handling of signed file; (3) The ability to understand how the evasive malware attack vectors work and record sighted attacks; (4) If CDR is not aware of a specific attack vector in the file, the attack vector may work. 

Therefore, CDR is not a silver bullet. However, it can still reduce the risk significantly since it does not rely on antivirus signatures or Artificial Intelligence (AI) detection \cite{tong2019improving}, that malware can be bypassed and mutate \cite{cycle}, and depends on disarming of generic attack vectors that exist in the files. Moreover, CDR is unsuitable for disarming code files, e.g., executable and scripts, since disarming potential attack vectors will break the software's ability to operate, unlike document files and multimedia files that could be safely disarmed. CDR should be developed per supported type since each file type has a specific file structure and different attack vectors that can be exploited.

In this paper, we propose a zero-trust Image Content Disarm and Reconstruction (ICDR) method to sanitize/disarm images while preserving the image functionality and quality as high as possible. Our method focuses on the Joint Photographic Experts Group (JPEG) image file type, which is the most common image format\footnote{https://1stwebdesigner.com/image-file-types/} \cite{aviad2020maljpeg}. The paper aims to validate ICDR prevention against JPEG malware dataset (from VirusTotal \cite{VirusTotal}) and verify against a second dataset (steganography data hiding attacks) where the second one is benign JPEG files that have been modified using steganography attacks \cite{opensteg,DCT-Image-Steganography,lsbsteg}. In addition, we release ICDR source code\footnote{https://github.com/ArielCyber/ICDR} and dataset creation tools for the community. Furthermore, for the first time, we validate the image file usability and the effect of the disarm and reconstruction steps on image quality and functionality. One of the main novelties of our work is that many works try to classify or identify that the image is malicious, but few try to disable the malicious content and reconstruct a new image.

The remainder of this paper is structured as follows: Section~\ref{Background} presented the standard image file format in general and JPEG in particular. Section~\ref{Related_work} surveys related work on the steganography concept and how to detection of malicious code in an image file; Section~\ref{Methodology} presents our ICDR system architecture while Section~\ref{Evaluation} describes the datasets that were used for this paper and the evaluation methods and metrics. The results of ICDR along with the quality impact on the dataset presents in Section~\ref{Results} while, Section~\ref{Conclusion} concludes and summarizes this paper.

\section{Background}
\label{Background}
Standard image files contain:
\begin{itemize}
    \item Image header: contains the image type, file type version, offsets to other headers and, the file's size. The header also contains content metadata, such as the date of file creation, location of the image and, the image owner's information.
    \item Metadata: contains additional image information such as height and width in pixels, type of compression used to store the file and, camera settings.
    \item Image pixel data: The color of a single pixel is represented as a series of three numbers between 0 and 255. These values correspond to Red, Green, and Blue (RGB). Overall, an RGB array represents the image pixels' color values.
\end{itemize}
The JPEG image filetype, which is the most common image format, can be found in the JPEG File Interchange Format (JFIF) specification \cite{Ham92}. A JPEG image file is a binary file which consists of a sequence of segments (can be contained in other segments hierarchically). Each segment starts with a two-byte indicator called a "marker" which helps to divide the file into different segments. A marker's start byte is 0xFF, and the second byte may have any value beside 0x00 and 0xFF. The marker indicates the type of data stored in the segment. All other markers are followed by the information about the size of the segment (two-byte integer) and the payload data contained in the segment. In order to embed malicious payload in JPEG images, one may try to find vulnerabilities in the viewer/parser program, while the other may try to use Steganography. Steganography is a technique used to hide information which can be text, another image, or malicious code inside the main image without affecting its quality, and the hidden information is invisible to the human eye and can be used for delivering pieces of code into the victim's host. Recently we have seen an increase in malware steganography hidden inside images, audio, and video \cite{steg-video}. The most recent attack by the Worok attack group uses to hide the next stage of their attack using a malicious PNG file that uses steganography to hide PowerShell script \cite{png-steg}. However, hiding the code inside the image is not the only way. Attackers use the image metadata to hide malicious scripts in JPEG images. For example, \cite{jpeg-steg} (md5: a96a07f25f14021f43312850ebe2b315) hides in the Artist and ImageDescription metadata tags a PHP code getting content from a URL that can also be used as a communication channel with the malware server. Another attack option is to hide the malware at the end of the file. In this method, the malicious payload is added after the EOF, the image is not affected, and it is easy to generate new attacks. However, hiding malware in metadata and at the end of a file is more evident in security detection.

Therefore it is possible to hide malicious code in an image without being noticed, by a viewer simply opening the image, since this will not affect the image quality or metadata besides the overall size of the file \cite{Karlo2021malware,dong2020ImageDetox,wiseman2017content}. This kind of payload can contain any form of a binary, including portable executable, Dynamic linking library (DLL), and Executable and Linkable Format (ELF) \cite{dong2020ImageDetox}. Furthermore, one may distinguish between a malicious payload carried by the JPEG images (e.g., in the payload of the file or the file headers) and JPEG images that carry hidden information (maybe even a malicious code piece) using steganography. In this research, we tackle both cases.

\section{Related work}
\label{Related_work}
Malicious code is usually hidden in one or a combination of the mentioned image file structure segments \cite{Karlo2021malware,dong2020ImageDetox,wiseman2017content,aviad2020maljpeg, DBLP:journals/corr/abs-2110-02504}. An attacker can hide scripts (e.g., JavaScripts, PHP) inside the image headers and/or inside the file's metadata that can lay dormant until another program activates it.

The metadata and scripts may contain abnormal values, such as abnormal height and width values, exploiting vulnerabilities in the program that is processing the file. For example, an exploit in Windows Media Player, when processing a malicious PNG file with abnormal image height and width \cite{MS05009}. These values can cause the media player to overload data from the file and load malicious code from the image payload, causing a buffer overflow. Puchalski et al. \cite{10.1145/3407023.3409187}
presented an algorithmic approach to detecting files with
unwanted hidden information appended to the GIF files. The detection focus on unwanted data added at the end of a file is shown to be relatively easy from an algorithmic viewpoint.

The steganography process involves hiding the data, usually encrypted, in the pixel data of the image \cite{osama2019hiding}. The simplest method is to hide the data in the Least Significant Bits (LSB) of a pixel value \cite{dong2020ImageDetox}. More modern algorithms, like Jsteg \cite{Rajesh2016Asecure}, allow the reduction of noise from the LSB steganography method, so it can be even harder to tell if a file holds hidden data. More robust methods \cite{wei2021secure,mason2020python} use various ratios of the pixel data, for example, Discrete Cosine Transform (DCT), which allows the payload to resist corruption by compression. In addition to hiding malicious scripts, steganography can also be used to embed sensitive user data (e.g., corporate information, personal information) inside images. An attacker can use the embedded images to leak the data by uploading the files to the attacker's website or a social media platform for later extraction without looking suspicious \cite{mamta2016review}.
Another significant advantage of steganography is the difficulty of detecting hidden data inside an image. Trying to detect steganography without knowing what type of steganography is used, is still a challenge for modern software \cite{aviad2020maljpeg,Rajesh2016Asecure,nandhini2020image,xiang2008review}. A detailed review of the image steganography techniques, including the Deep Learning (DL) models and the recent Generative Adversarial Networks (GAN) based models for hiding data, can be found in \cite{DBLP:journals/corr/abs-2110-02504}. Although vast research about steganography techniques exists, detection of malicious JPEG images (images that contain malware or malicious code, which are found in the file headers, end of file or hidden in the RGB)  based on machine learning approaches are not common \cite{aviad2020maljpeg}. One of the main reasons could be the lack of available datasets for classifying the malware \cite{DBLP:journals/corr/abs-2110-02504}. Dubin \cite{dubin_CDR} presented the first CDR system to RTF file and showed that CDR was able to prevent 100\% of the attacks while the CDR output file is highly similar to the original document.

The detection of malicious code in an image depends on the vector of an attack and its complexity. For example, payloads can be detected easily by examining the file size or additional data after the EOF bit. On the other hand, detecting malicious code encrypted inside the file requires an analysis of every segment of the image file structure \cite{wiseman2017content}. Therefore, detecting them requires an extensive search for anomalies in the image structure that can be interpreted as possible encrypted data. 
Most of the models presented are focusing on detection and not prevention. In this paper, we present a prevention method based on disarming evasive attack vectors and reconstructing the image as a new image regardless of whether it is detected as malicious. Our system achieves these goals with minimum effect on quality.

\section{Methodology}
\label{Methodology}
As stated before, our paper aims to develop a system to disarm image files from possible embedded malicious code, and hidden messages (steganography data) while keeping high image quality.

\begin{figure}[h!]
\centering
\includegraphics[width=0.7\textwidth]{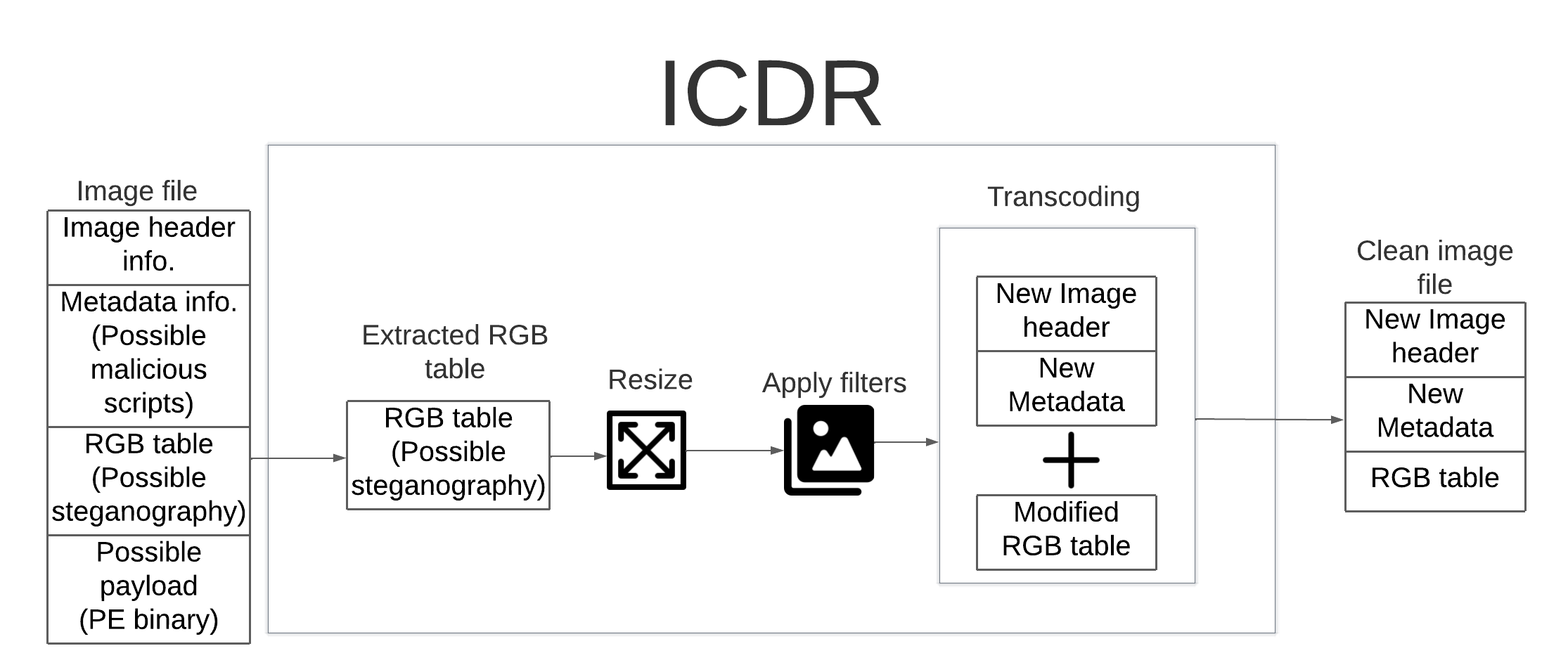}
\caption{ICDR system}
\label{fig:ICDR system}
\end{figure}

\begin{figure}[h!]
\centering
\includegraphics[width=0.7\textwidth]{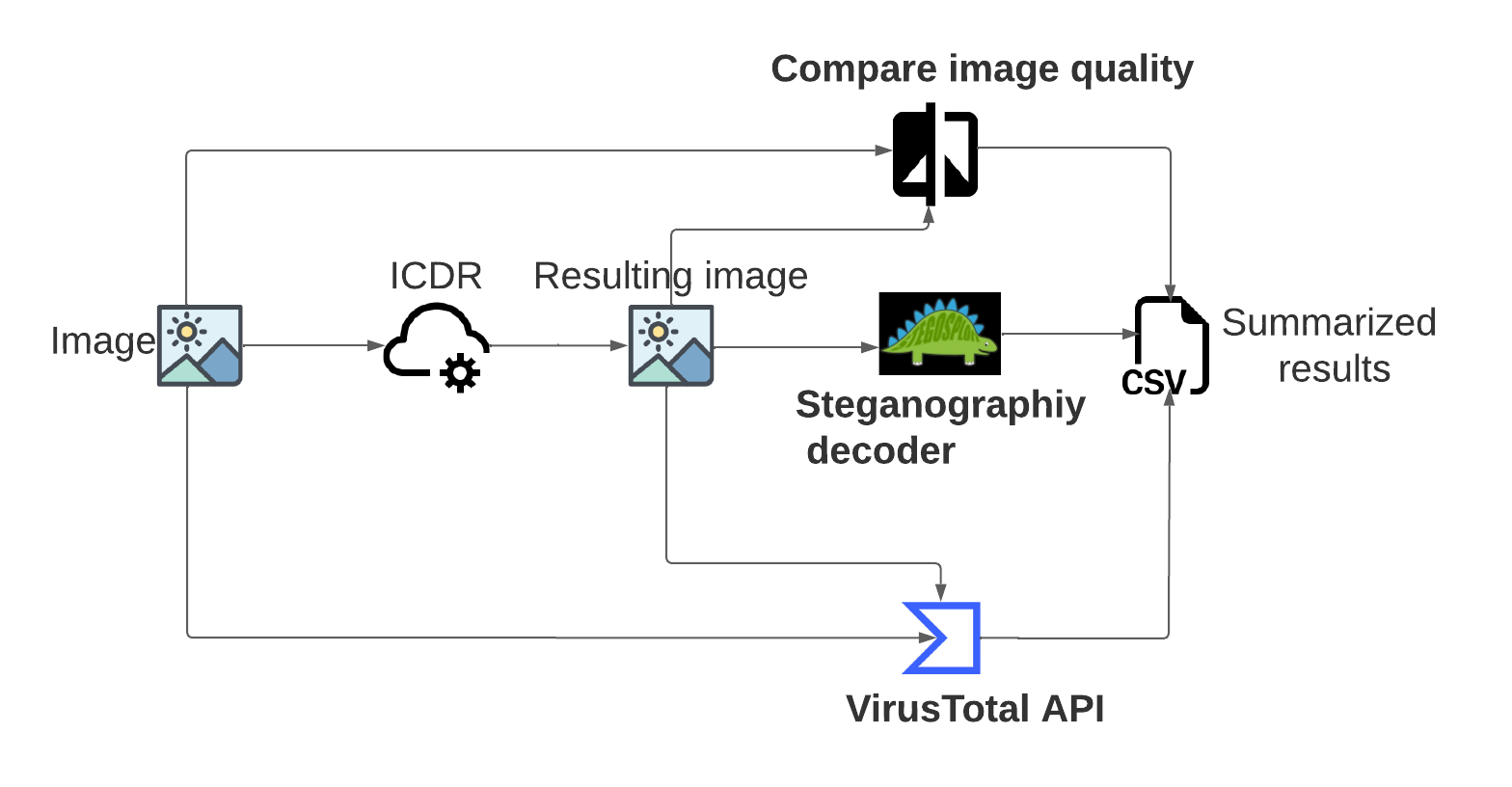}
\caption{ICDR -  Evaluation Process}
\label{fig:evaluation}
\end{figure}

Our ICDR system architecture is illustrated in Fig. \ref{fig:ICDR system} and takes a JPEG file as an input (the file may be benign or malicious). The ICDR system extracts its RGB table/content from the input image file and builds a new image based on the RGB. The extraction automatically removes all image metadata, malicious content hidden in the image file header, and hidden content at the end of the file. However, advanced malware and data hiding methods use the image content to evade detection. Therefore, ICDR suggests an additional three image disarm steps: image resize, image filters, and transcoding. 
Because some malicious files do not follow the structure of an image file (PHP script with JPEG magic bytes) or have an unreasonable image size (usually to overflow the system), our system checks if the number of image pixels is between one and thirty million, and if the files are detected to be valid by the Aspose Imaging library \cite{aspose}. Any image outside the range is considered corrupt by the ICDR system.  For the evaluation phase, as can be seen from Figure \ref{fig:evaluation}, we send in parallel the reconstructed file to VirusTotal and to the steganography decoder to validate our results, and both results are reported into a CSV file. We also validate the image quality between the original and the reconstructed ICDR image. Please note that ICDR uses VirusTotal and the steganography decoder as the success criteria. In the following we describe in detail each ICDR step. Note that to implement the methods, we utilize Aspose imaging \cite{aspose} functions to edit the inputted images.

\subsection{Image resize}
Resizing images is a process of changing the image's size and scale of the RGB table to the desired size while maintaining the image, for example, changing an image of the size 256 X 256 pixels to 512 X 1048 pixels or to 128 X 128 pixels.
When decreasing the size of an image, there is a need to discard several pixels and replace them with one pixel with RGB values that can represent the discarded pixels to maintain the image with the new scale according to the resize algorithm.
When increasing the image size, the resize algorithm spreads the current pixels to a new RGB array with the given size. To fill in the missing pixels, usually, the algorithms create new pixels based on the original neighboring pixels' values and the distance from them. In this work, we used the well-known resize algorithm, bilinear image resizing \cite{Bilinear2018}.

In the resize method, we change the size of the image to a 0.97 scale and back to the original size.
The resize method allows for removing 3 percent of pixels and recreating them using the bilinear resize algorithm. 

\subsection{Filters}
An image filter is a technique through which an image's size, colors, shading, and other characteristics are altered. An image filter is used to transform the image using different graphical editing techniques. Typically, the image filter graphical editing techniques include options such as: editing the color scheme/theme/contrast of the image; adjusting image brightness; adding effects to the image; changing the texture \cite{filterDefinition}. By applying filters on steganography-infected images, it is possible to corrupt the flow of data inside the color values of pixels and the ratios they create. To that end, we used existing filters \cite{SharpeningDefebition,GBFdefebition,medianDefinition,GWFdefinition,bilateral2008sylvain} to corrupt potential embedded malicious code inside the image. However, unlike the general use of filters, we maintain the filter output, so the distortion created by the filters will be minimal.

By changing the contrast, brightness, and texture of the image this alters the RGB color values and the coefficients between neighboring pixels. This kind of change allows us to damage most embedded data inside the image \cite{zhiying2021destroying}. 
For our ICDR system, we used Gaussian blur that was found to be the most effective by followed by Sharpen filter on the inputted image, we found that combination provided the best results (the results can be see in the third table in  https://github.com/ArielCyber/ICDR). 

\subsection{Transcoding}
Transcoding is the process of converting a file from one encoding format to another compatible file encoding format. For example, converting a JPEG file to PNG, WMF, or PDF. 

By changing the file type, we also change the file structure, which can make the file incompatible for the attacker to decode the malicious code \cite{yiyi2016image}. In our ICDR system, we do not change the original file directly using transcoding but create a new file of a different format and back to the original format (JPEG to PNG to JPEG). By using the original image RGB table, we create a new file according to the given type with a RGB table data (that is copied from the original file), discarding any data outside the RGB table and its scale.

This way, we remove any metadata, additional information in the file's headers, and any payload if the payload exists inside the file. 
The problem with transcoding to different file types is that a drastic change to the file type can reduce its usability for the user, as a PDF cannot always be a good replacement for a JPEG.
That is why in our ICDR we transcode the new file to the original format.
Furthermore,
transcoding images can deal with steganography if the resulting transcoded file type implements a different encoding for the pixel data. For example, when transcoding a JPEG file to a PNG file, the difference in the encoding allows us to damage the hidden data inside the file's pixel data.
However, this method relies on the differences of encoding between file formats, as when
transcoding a steganography infected JPEG file to SVG, and then to JPEG, the steganographic decoder is able to read the hidden data inside the new JPEG file.

\section{Evaluation}
\label{Evaluation}
In this section, we describe the evaluation of our ICDR method. First, we describe the dataset, then our methods to deploy steganography attacks and decoders are presented, and finally, our quality metrics are presented. 

\subsection{Dataset}
The data-set that was used for this research is 1805 infected JPEG images from VirusTotal and 3283 benign JPEG images with varied quality and sizes, taken from Kaggle's Prasun Roy natural-images dataset \cite{prasun2018natural}. The VirusTotal files are made up of JPEG images containing embedded scripts, payloads, and exploits inside the images (details can be found at Table \ref{vt_dataset}). The steganography tools that have attacked each benign file are \cite{opensteg,lsbsteg,DCT-Image-Steganography} as described in Section \ref{Evaluation methods}.

\begin{table}
\begin{center}
\begin{tabular}{ | p{5cm} | p{3.1cm}| } 
\hline
     & \textbf{Percentage}    \\ 
  \hline
    \textbf{Iframe} & 36\%  \\ 
  \hline
    \textbf{PHP} & 26\%  \\ 
  \hline
    \textbf{Trojan} & 9\%  \\ 
  \hline
    \textbf{JavaSscript} & 6\%  \\ 
  \hline
    \textbf{Exploits} & 6\%  \\ 
  \hline
    \textbf{HTML} & 4\%  \\ 
  \hline
    \textbf{Script} & 4\%  \\ 
  \hline
    \textbf{Backdoor} & 3\%  \\ 
  \hline
    \textbf{Virus} & 2\%  \\ 
  \hline
    \textbf{Miner} & 2\%  \\ 
  \hline
    \textbf{Android} & 1\%  \\ 
  \hline
    \textbf{Ransomeware} & 1\%  \\ 
  \hline
\end{tabular}
\caption{Threats composition of VirusTotal dataset as seen in \cite{aviad2020maljpeg}}
\label{vt_dataset}
\end{center}
\end{table}

\subsection{Evaluation methods}
\label{Evaluation methods}
As stated in the methodology section, to evaluate the ICDR system, we check the existence of malicious code or embedded steganographic data inside the image using Application Programming Interface(API) of VirusTotal and appropriate steganography decoder over the input and the resulting files. VirusTotal validates the system's success rate in removing potential malware. To simulate steganographic attack success, we embed messages inside the benign data set and try to decode the message back. We used the following steganography attacks:
\begin{itemize}
    \item OpenSteg \cite{opensteg}: A tool that hides data in the Least Significant Bits (LSBs) of the RGB values. It breaks the payload into bits and changes the LSB in each color according to the payload \cite{devang2018LSB}
    \item DCT-Image-Steganography \cite{DCT-Image-Steganography}: This tool utilizes an image's  Discrete Cosine Transform (DCT) values to hide data. The tool calculates the DCT value table and encodes the payload inside the table's values \cite{osama2019hiding,Rajesh2016Asecure}
    \item A modified LsbSteg \cite{lsbsteg}: The tool also uses the LSB method of data hiding; however, we modified the tool to hide the data in the Most Significant Bit (MSB) of the RGB values, making the encoded message visible in the image. Unlike other steganographic methods that try to hide data as much as possible, this method does the opposite, therefore requiring a more drastic change to the image. 
\end{itemize}

To evaluate the ICDR results, we decided to compare our ICDR to a state-of-the-art approach to remove steganographic data in images named Detox \cite{dong2020ImageDetox}. The Detox method uses a nonlinear transfer function to remove hidden data. The method applies the function $ New[RGB] $ = $ W \cdot [RGB] ^ {1/\gamma} $ on every RGB pixel value, where $\gamma$ denotes a value within a specific range ($0.950 < \gamma < 0.995$ or $1.005 < \gamma < 1.050$) characterizing a nonlinear transfer function, and $W$ represents the application of an alpha channel. The function works as a filter on the image, darkening the brightest areas while leaving most of the dark areas visibly unchanged. 

\subsection{Quality metrics}
We decided on preferred well-known quality metrics for comparing the quality effects of our ICDR system's methods. The used metrics are Universal Image Quality Index (UQI) \cite{zhou2002UQI}, Peak Signal-to-Noise Ratio (PSNR) and Structural SIMilarity (SSIM) \cite{zhou2004SSIM}. The value of the present metrics is measured by comparing an altered image to its original value and measuring the resulting differences created by the altering of the original image. It should be known that the metrics presented do not always align with the subjective image quality, as measuring using those metrics regards the original image as the optimal value even if the quality is poor. 
Universal Image Quality Index (UQI) is a method measuring distortions in luminance and contrast, in addition to the signals of the equated images. The UQI metric values are between -1 and 1, 1 being the best value \cite{zhou2002UQI}. Peak Signal to Noise Ratio (PSNR) measures the ratio between the maximum possible power of a signal and the power of corrupting noise that affects the quality of the image. PSNR is usually expressed as a logarithmic quantity using the decibel scale, meaning that the higher the PSNR value the better the image quality. PSNR is most commonly used to estimate the efficiency of compressors but can also be used to measure the effects of filters and other reconstructions on the image quality. Therefore PSNR is not a measure of subjective image quality but rather a measure of the efficiency of the image transformation \cite{zhou2004SSIM}. Structural similarity (SSIM), similarly to UQI, compares each signal value of the equated images and measures the differences in contrast and luminance. In addition, the metric also takes into account the structure of the objects in the image. The metric values are between 0 and 1, 1 being the best value \cite{zhou2004SSIM}.

First we wanted to show that the steganography tools that we are using (OpenSteg, MSB, DCT-Image-Steganography) have minimal effect on the image quality. The results as can be seen in Table \ref{steg quality effects} show minimal effect over each one of the quality metrics (UQI, PSNR and SSIM)  after we used the steganography tool. For example, the UQI value is between 0.99999 - 0.99911.
\begin{table}
\begin{center}
\begin{tabular}{ | l | c| c | c | } 
\hline
     & \textbf{UQI} & \textbf{PSNR} & \textbf{SSIM}  \\ 
  \hline
    \textbf{OpenSteg} & 0.99999 & 69.4909 & 0.99989 \\ 
  \hline
    \textbf{MSB} & 0.99991 & 39.6968 & 0.99782 \\ 
  \hline
    \textbf{DCT-Image-Steganography} & 0.99911 & 38.3715 & 0.97451 \\ 
  \hline
\end{tabular}
\end{center}
\caption{Quality values (UQI, PSNR, SSIM) of the Steganography tools (OpenSteg, MSB, DCT-Image-Steganography)}
\label{steg quality effects}
\end{table}
From the table we can see that there is minimal effect on the image quality so now we can continue to see the results of our ICDR novel system. Before discussing the results we presenting how we tune the ICDR parameters.

\subsection{ICDR Tuning}
In this section, we present how we tune the ICDR.
First, we wanted to find the right scale for the resizing method. Therefore, we tested several scales from 95\% to 99\% as can be seen in Table \ref{resize results} and ranked them according to the quality effect and the success rate of removing the steganographic data. From the results we decided to choose a resize of 97\% because of its success rate relative to quality.
\begin{table}
\begin{center}
\begin{tabular}{ | p{0.6cm} | p{1cm}| p{1cm} | p{1cm} | p{1.5cm} | } 
\hline
    & \textbf{UQI} & \textbf{PSNR} & \textbf{SSIM} & \textbf{Success rate}  \\ 
  \hline
    99\% & 0.99981 & 45.4035 & 0.99818 & 89.5\% \\ 
  \hline
    98\% & 0.99962 & 42.635 & 0.99674 & 96\% \\ 
  \hline
    97\% & 0.99945 & 41.1892 & 0.9954 & 100\% \\ 
  \hline
    96\% & 0.99929 & 40.003 & 0.99391 & 100\% \\ 
  \hline
    95\% & 0.9991 & 39.1864 & 0.99246 & 100\% \\ 
  \hline
\end{tabular}
\end{center}
\caption{Quality values (UQI, PSNR, SSIM) of the resize scale (99\% - 95\%) and the success rate (last column) to disable MSB steganography}
\label{resize results}
\end{table}

Another step in ICDR is the filter method. In order to distort the image while maintaining high quality, several layers of filters are applied to the inputted image. Therefore, to understand how to build our filter step, we tested up to six layers of filters. The filters used are: Gaussian blur (GBF) \cite{GBFdefebition}, Bilateral smoothing (BSF) \cite{bilateral2008sylvain}, Gauss Wiener (GWF) \cite{GWFdefinition}, Median filter (MF) \cite{medianDefinition}, and Sharpen filter (SF) \cite{SharpeningDefebition}. From the results as can be seen in Table \ref{filter_ranks} we came to the conclusion that the best combination of filter layers is the Gaussian blur (GBF), followed by Sharpen filter (SF). 
\begin{table}
\begin{center}
\begin{tabular}{ | p{3cm} | p{1.4cm}| p{1.4cm} | p{1.4cm} | } 
\hline
     & \textbf{UQI} & \textbf{PSNR} & \textbf{SSIM}  \\ 
  \hline
    \textbf{GBF, SF} & 0.99925 & 35.3191 & 0.96782 \\ 
  \hline
    \textbf{BSF, SF} & 0.99914 & 35.0188 & 0.96102 \\ 
  \hline
    \textbf{SF, GBF} & 0.99916 & 34.6787 & 0.9656 \\ 
  \hline
    \textbf{GBF, SF, SF, GBF} & 0.99888 & 33.7073 & 0.96066 \\ 
  \hline
    \textbf{GBF, SF, GBF, SF} & 0.99887 & 33.6459 & 0.96063 \\ 
  \hline
    \textbf{SF, BSF, GBF, SF} & 0.99878 & 33.4952 & 0.95587 \\ 
  \hline
    \textbf{SF, BSF, SF, GBF} & 0.99875 & 33.3851 & 0.95558 \\ 
  \hline
    \textbf{SF, GBF, SF, GBF} & 0.9988 & 33.3199 & 0.95874 \\ 
  \hline
    \textbf{BSF, SF, GBF, SF} & 0.99874 & 33.3124 & 0.95423 \\ 
  \hline
    \textbf{SF, GBF, GBF, SF} & 0.99879 & 33.2898 & 0.95858 \\ 
  \hline
\end{tabular}
\end{center}
\caption{Filter layers quality values (ranking by PSNR values)}
\label{filter_ranks}
\end{table}

\section{Results}
\label{Results}
In this section, we will review the results of ICDR along with the quality impact on the dataset. We compared each prevention method as a standalone system and combined as our ICDR. First, we tested ICDR against the malicious image file from Virustotal. Then, we took the benign images from Kaggle and used the stag-tools to attack the images and see if our ICDR could disarm and reconstruct them. 
\begin{table}[h!]
\begin{tabular}{|p{2cm}|l|l|l|l|l|l|}
\hline
steg-attack & Dataset & Resize & Filters & Tr-code & Detox\cite{dong2020ImageDetox}  & ICDR  \\
\hline
-           & VT      & 0\%    & 0\%     & 100\%     & 0\%    & 100\% \\\hline
OpenSteg    & Kaggle  & 100\%  & 100\%   & 100\%       & 100\%  & 100\% \\\hline
DCT         & Kaggle  & 100\%  & 100\%   & 100\%       & 100\%  & 100\% \\\hline
MSB         & Kaggle  & 100\%  & 100\%   & 100\%       & 85.7\% & 100\% \\\hline
\end{tabular}
\caption{Success rate of each part of the ICDR as standalone, Detox and ICDR over dataset (VT, Kaggle) under steganography attack (OpenSteg, DCT, MSB). Note that, Tr-code is for transcoding}
\label{tab:results}
\end{table}
\begin{table}[!]
\begin{center}
\begin{tabular}{ | c | c|c | c| } 
\hline
     & \textbf{UQI} & \textbf{PSNR} & \textbf{SSIM}  \\ 
  \hline
    \textbf{Resize} & 0.99945 & 41.1892 & 0.9954 \\ 
  \hline
    \textbf{Transcode} & 0.99866 & 37.5694 & 0.96888 \\ 
  \hline
    \textbf{Filters} & 0.99819 & 36.9803 & 0.97541 \\ 
  \hline
  \textbf{Detox \cite{dong2020ImageDetox}} & 0.99827 & 33.5694 & 0.99594 \\
  \hline
    \textbf{ICDR} & 0.99644 & 32.9294 & 0.96274 \\ 
  \hline
\end{tabular}
\caption{Average Quality Impact}
\label{quality impact ICDR}
\end{center}
\end{table}
\begin{table}[!]
\begin{center}
\begin{tabular}{ | p{0.9 cm} | c | c | p{0.75 cm} | c | c | }   
\hline
      & \textbf{Resize} &  \textbf{Filters} & \textbf{Trans-code} & \textbf{Detox \cite{dong2020ImageDetox}} & \textbf{ICDR} \\ 
  \hline
    \textbf{Success rate}  & 0\% & 100\% & 100\% & 100\% & 100\% \\ 
  \hline
\end{tabular}
\caption{ICDR over the anti-resize steganography}
\label{anti_resize_rates}
\end{center}
\end{table}
\subsection{Embedded Malware}
As a pre-processing step, we filtered images by total pixel number and the Aspose imaging \cite{aspose} library's ability to read the files. Those, files whose total pixel number exceeds 30,000,000 or are unreadable by the library are considered corrupt or malicious.
Therefore, when the images were subjected to ICDR, we were able to process only 88.2\% of the total images (1592 out of 1805); the rest were labeled corrupt. We took each malicious file as the input file for the ICDR. The input and result files sent to VirusTotal for validation. Out of the processed files (1592), we achieved 100\% of malware removal. This means that our ICDR system could disarm all the malicious files. One of the main reasons for the success of our system is that our methods do not change the actual file but create a new one by copying the RGB array and creating the appropriate image header and metadata containing the new file attributes. Because the malware is embedded outside the RGB array, the new file will not contain the embedded malware, making it safe to use. 

Table \ref{tab:results} presents detailed results where not only did we test the ICDR but also each step of it as standalone and comparison to the state-of-the-art sanitizing method (Detox \cite{dong2020ImageDetox}). Note that the results (first line) are based on evaluation with VirusTotal API. 
The resize standalone results are poor due to the fact that the manipulation image (RGB values) will not effect the embedded malware, leaving the file as a threat. However, using the entire ICDR methods it is possible to remove malware hidden outside the RGB array, from the ICDR results. Note that the part of helping ICDR to disarm the malicious part in this case is the transcoding.

\subsection{Steganography Attacks}
In the following Table \ref{tab:results} (OpenSteg, DCT, MSb rows), we use the image file from Kaggle, 3282 JPEG benign files. In each test we take the files and use one steganography tool:  OpenSteg \cite{opensteg}, DCT \cite{DCT-Image-Steganography}, and MSB tools. After the attack we want to evaluate our ICDR and compare it with Detox \cite{dong2020ImageDetox}. Therefore, we evaluate each of the ICDR steps as a standalone solution and, the entire ICDR system.
From the results of Table \ref{tab:results} (in the OpenSteg row), we can see that the ICDR steps (resize, filter, transcode) disarm the malicious as standalone and the ICDR gets the advantages of these three steps. The Detox has the same results as ICDR. The same results can be seen in the case of DCT attack (Table \ref{tab:results}). From the results of MSB (Table \ref{tab:results}) we can see that ICDR outperforms Detox.  



\subsection{Quality}
To evaluate the ICDR, we should also evaluate the image quality before and after the ICDR. We measure the average quality results by comparing the resulting images to the original as the optimal version, using the metrics of UQI, PSNR, and SSIM, as can be seen in Table \ref{quality impact ICDR}. In the case of transcoding, because the RGB values are being copied to a new file the resulting image quality is the same as the original. From the results we can see that as each part of the ICDR is standalone, the entire ICDR and Detox didn't dramatically decrease the quality.

\subsection{Discussion}
One may conclude that due to the fact that standalone ICDR steps achieved 100\%, there is not advantage in connecting them together to one methodology (ICDR). However, based on our experiments while using steganography and trying to use transcoding as a standalone solution, there are some cases (i.e., jpeg-to-jpeg transcoding) where transcoding did not remove the hidden data. Moreover, due to the fact that we cannot know where the malicious code is, it is essential to use the entire system.
Therefore, we can conclude that ICDR can provide the highest level of protection. Any ICDR system should be robust and deal with not only current attacks but can also be used against the next step attacks (sophisticated attacks). To evaluate the robustness of our system, we develop a small steganographic tool that can decrease the resize detection results. The steganographic tool counteracts most resizing of images while maintaining the embedded data \cite{antiresize}.
By creating a large image (4000 X 4000 pixels) and splitting it into 8 X 8 subsections and embedding every pixel of the subsections with our message, we were able to maintain the embedded data even after resizing the image to half size, or when widening and thinning the image.
Because of the way resizing algorithms rely on neighboring pixels' values to create or remove pixels to match the right scale, creating chunks of the same color value can maintain the color with its embedded message intact. The results can be seen in Table \ref{anti_resize_rates}.

Malware can be hidden in the meta data of a file, at the end of a file, or even in a picture. As this paper shows, there is no 100\% detection in the case of pictures, and it is better to use ICDR to avoid vector attacks instead of taking risks. Using ICDR, in one way, presents a 100\% prevention, while negligible image quality is achieved. Moreover. based on the assumption that changing the picture format is feasible with a low level of computation, computation and energy are a few milliseconds per file.

\section{Conclusion}
\label{Conclusion}
In this study we present a novel Image Content Disarm and Reconstruction. Based on two datasets, we show that our novel ICDR can disarm malicious JPEG files in both cases (malicious in the pixels and outside the pixels, such as malicious code in the file headers and hidden in the end of the file). Moreover, we also show the performance of our new system against new attacks. 
As for future work, we plan to extend ICDR to support and validate additional image file types such as PNG, DICOM, SVG, and others. To do that, we plan to extend the malicious JPEG malware dataset and extend it to additional malware-infected files from those formats and research steganography attacks matching those file formats.
\section*{Acknowledgment}
The authors would like to thank VirusTotal for granting us access to their cloud service and malware collection for educational use and Aspose for providing an academic license for their ASPOSE total product libraries for parsing and manipulating image files. This work was supported by the Ariel Cyber Innovation Center and the Israel National Cyber Directorate of the Prime Minister’s Office.
\newpage

\bibliographystyle{unsrt}
\bibliography{main.bib}

\end{document}